\documentclass[twocolumn,superscriptaddress,showpacs,aps,prb]{revtex4}
\usepackage{makeidx}
\usepackage{bm}
\usepackage{graphics}
\usepackage[dvips]{epsfig}
\usepackage{mathtext}
\usepackage{amsmath}
\newcounter{fig}

\begin{document}
\title{Optical properties of   graphene }
\author{L.A. Falkovsky}
\address{L.D. Landau Institute for Theoretical Physics, Moscow
117334, Russia} \address{Institute of the High Pressure Physics,
Troitsk 142190, Russia}
\begin{abstract}
Reflectance and transmittance of graphene in the optical region
are analyzed as a function of frequency, temperature, and carrier
density. We show that the optical graphene properties are
determined by the direct interband electron transitions. The real
part of the dynamic conductivity in doped graphene at low
temperatures takes the universal constant value, whereas the
imaginary part is logarithmically divergent at the threshold of
interband transitions.
\end{abstract}

\maketitle
\section{Introduction}
Monolayer and bilayer graphenes  are gapless two-dimensional (2D)
semiconductors  whereas its 3D predecessor, graphite, is a
semimetal. Hence the dimensionality effects can be studied for the
unique substance. Monolayer graphene has a very simple electron
band structure. Near the energy $\varepsilon=0$, the energy bands
are cones $\varepsilon_{1,2}(\mathbf{p})=\pm vp$ at the $ K$
points in the 2D Brillouin zone with the constant velocity
parameter $v=10^8$ cm/s.  Such a degeneration is conditioned by
symmetry because the small group $C_{3v}$  of the $K$ points has a
two-dimensional representation.

While the carrier concentration is decreasing in the field gate
 experiment, the graphene conductivity at low temperatures
 goes to the finite minimal values.
Much theoretical efforts  have been devoted to evaluate the
minimal conductivity in different approaches. Theoretical  and
experimental researches  show that the main mechanism of the
carrier relaxation is provided
 by the charged impurities and gives the collision  rate
  $\tau^{-1}\sim 2\pi^2e^4n_{imp}/\hbar\epsilon_g^2\varepsilon$, where
 $\epsilon_g$ is the dielectric constant of graphene, $\varepsilon$ is
 the characteristic electron energy (of the order of  the Fermi energy
 or temperature),
 and $n_{imp}$ is the density of charged impurities per the unit surface.
 The general quantum expression for the  conductivity depending on the
 frequency $\omega$ and wave vector $k$ has been derived in our paper \cite{FV}.
 In   various limiting cases,
 our result coincides  with the formulas  of Refs. \cite{GSC,Cs}.
This expression is valid under a restriction that the collision
rate of carriers is less than the frequency and spatial
dispersion, $\tau^{-1}\ll\omega, kv $.
  In the optical range, we can neglect the spatial dispersion of conductivity
 compared with the frequency dependence and use the dynamical
 conductivity $\sigma(\omega)$ to study the graphene optical properties \cite{FP}.
 The optical visibility of both monolayer and bilayer graphene was
 theoretically   studied in Ref. \cite{ARF} focusing on the  role of the
    underlying substrate. Recently, graphene transmittance spectra
    were observed \cite{NBG} and the dynamical conductivity was found
    as $\sigma(\omega)=e^2/4\hbar$ in agreement with the theory \cite{FV,GSC,Cs}.

In the present paper, we analyze in detail the optical properties
of the graphene monolayer and multilayers.  We use the dynamic
conductivity $\sigma$ as a function of frequency $\omega$,
temperature $T$, and chemical potential $\mu$. The chemical
potential of ideal pure graphene at any temperature is at the
crossing of the bands $\varepsilon_{1,2}$. With the help of the
gate voltage, one can control the density and type ($n$ or $p$) of
carriers varying their chemical potential. Using the dynamic
conductivity and the appropriate boundary conditions at
interfaces, we calculate the reflection and transmission
coefficients of graphene monolayer and multilayers with varying
carrier concentration.
\section{Optical conductivity of graphene}
The general expression for the conductivity used here is obtained
in the previous paper \cite{FV}. For high frequencies,  $\omega\gg
(kv, \tau^{-1})$, the dynamical conductivity [see  Eq. (8) in Ref.
\cite{FV}] is given by
\begin{equation}
  \sigma(\omega) = \frac{e^2\omega}{i\pi\hbar}\left
  [\int\limits_{-\infty}^{+\infty} d\varepsilon\frac{|\varepsilon |}{\omega ^2}
   \frac{df_0 (\varepsilon)}{d\varepsilon}- \int\limits_0^{+\infty}
  d\varepsilon\frac{f_0 (-\varepsilon)-f_0(\varepsilon)}{(\omega+i\delta)^2 -
  4\varepsilon^2}\right]\, ,
  \label{sigma}
\end{equation}
where $f_0(\varepsilon)=\{\exp[(\varepsilon-\mu)/T]+1\}^{-1}$ is
the Fermi function.

The first term in Eq. (\ref{sigma}) corresponds to the intraband
electron--photon scattering processes.  Integrating, we obtain
explicitly:
\begin{equation}
     \sigma ^{intra}(\omega) =\frac{2ie^2T}
     {\pi\hbar(\omega+i\tau^{-1})}
\ln{[2\cosh(\mu /2T)]}
 \label{sigm}    \, .
 \end{equation}
Here, we write   $\omega+i\tau^{-1}$  instead of $\omega$ in order
to take into account the electron -- impurity scattering
processes. In such a form the intraband conductivity coincides
with the Boltzmann-Drude expression. For the Fermi-Dirac
statistics, $\mu\gg T$, the intraband conductivity takes the form
\begin{equation}
     \sigma ^{intra}(\omega) =\frac{ie^2|\mu|}
     {\pi\hbar(\omega+i\tau^{-1})}
 \label{si}    \, ,
 \end{equation}
where the chemical potential determines the carrier concentration
$n_0=(\mu/\hbar v)^2/\pi$.

The second term in Eq. (\ref{sigma}), where $\delta\to 0$ is
infinitesimal quantity determining the bypass around the integrand
pole, owes its origin to the direct interband electron
transitions. The integral is easily evaluated in the case of zero
temperature:
\begin{equation}
    \sigma^{inter}(\omega) =
    \frac{e^2}{4\hbar}\left[\theta(\omega-2\mu)-\frac{i}{2\pi}\ln
    \frac{(\omega+2\mu)^2}{(\omega-2\mu)^2}
     \right]\, .
 \label{ibd} \end{equation}
Here and below, we consider for simplicity the case of the
positive $\mu$. The step function $\theta(\omega-2\mu)$ conveys
the condition for the interband electron absorption at low
temperatures. The expression shows that the interband contribution
plays the leading role around the absorption threshold
$\omega\approx2\mu$ comparatively with  the Boltzmann-Drude
intraband term, Eqs. (\ref{sigm}) and (\ref{si}), which is
important at relatively low frequencies, $\omega<\mu$. The
logarithmic singularity is cut off with temperature.  At the
finite, but low temperatures, the following substitutions should
be made in Eq. (\ref{ibd})
\begin{eqnarray}
\theta(\omega-2\mu)\rightarrow\frac{1}{2}+\frac{1}{\pi}\arctan[(\omega-2\mu)/2T]\\
\nonumber (\omega-2\mu)^2\rightarrow(\omega-2\mu)^2+(2T)^2\,.
\label{sub}\end{eqnarray} The electron relaxation processes
produce  similar smearing.

 It is useful for  numerical calculations to present the
  difference of the Fermi functions in the second integrand  (\ref{sigma}) as
 $$G(\varepsilon)=\frac{\sinh(\varepsilon/T)}{\cosh(\mu/T)+\cosh(\varepsilon/T)}\, .$$
Adding and subtracting $G(\omega/2)$ in the numerator of the
integrand, and noticing that the   the principal value of the
integral with $G(\omega/2)$ equals to zero, we arrive at the
integral without singularities. Then we can write the interband
conductivity in the form available for numerical calculations:
  \begin{equation}
    \sigma^{inter}(\omega) =
    \frac{e^2}{4\hbar}\left[G(\omega/2)
     -\frac{4\omega}{i\pi}\int\limits_0^{+\infty}
  d\varepsilon\frac{G(\varepsilon)-G(\omega/2)}{\omega^2 -
  4\varepsilon^2}\right]\, .
 \label{equat6} \end{equation}
For the Fermi-Dirac and Boltzmann   carrier statistics,
correspondingly, the first term is given asymptotically by
\begin{eqnarray}
G(\omega/2)= \left\{
\begin{array}{ll}
 \displaystyle\theta(\omega-2\mu), & \mu\gg T\, , \\
\displaystyle\tanh(\omega/4T) , & \mu\ll T\, .
\end{array}
\right. \label{g} \end{eqnarray}
\begin{figure}[h]
\noindent\centering{
\includegraphics[width=90mm]{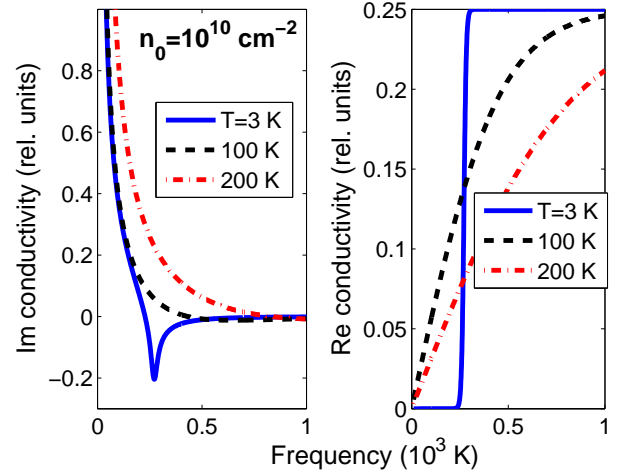}
} \caption{(color online). Imaginary (left)  and real (right)
parts of conductivity in units of $e^2/\hbar$ for graphene doped
with the carrier density $n_0=10^{10}$ cm$^{-2}$ at temperatures
noted at lines. The chemical potential equals to 135, 65, and 33 K
at 3, 100, and 200 K, correspondingly. } \label{uds}
\end{figure}
The conductivity calculated with the help of Eqs. (\ref{sigm}),
(\ref{equat6}) is shown in Fig. \ref{uds}. The step function
behavior of  the real part (absorption) and logarithmic
singularity of the imaginary part are clearly seen at low
temperatures. While increasing the  temperature, the transition
from the Boltzmann  statistics to the Fermi-Dirac statistics is
also evident.

\begin{figure}[h]
\noindent\centering{
\includegraphics[width=90mm]{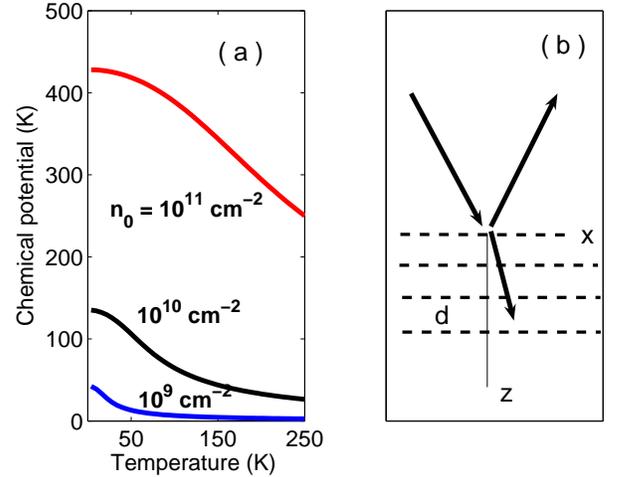}
} \caption{(color online). (a) Chemical potential (in K) as a
function of temperature at carrier densities noted at curves. (b)
Multilayers sample and geometry of wave scattering.} \label{mu}
\end{figure}

  By using the gate
voltage, one can control the density of electrons ($n_0$) or holes
($-n_0$). Then the chemical potential is determined by the
condition
\begin{eqnarray}
n_0 =\frac{2}{\pi(\hbar
v)^2}\int\limits_0^{+\infty}\varepsilon[f_0(\varepsilon-\mu)-
f_0(\varepsilon+\mu)] d\varepsilon\, . \label{den}\end{eqnarray}
From this expression and Fig. \ref{mu}(a), one can see that the
chemical potential
  goes to zero while the temperature increases.

\section{Spectroscopy of graphene layers}

In order to calculate the graphene reflectance, we apply Maxwell's
equations
\begin{equation}
\label{maxwell} \nabla (\nabla \cdot {\bf E}) -\nabla^2 {\bf
E}=\epsilon_0\frac{\omega^2}{c^2}{\bf E}+ \frac{4\pi
i\omega}{c^2}{\bf j}\, ,
\end{equation}
where $\epsilon_0$ is the ion contribution into the dielectric
constant and ${\bf j}$ is the conductivity current. We consider
the case of the $p$-polarization, when the field ${\bf E}$ lies in
the $xz$ plane and the current ${\bf j}$ has  only the in-layer
$x$ component (see Fig \ref{mu}b).

\subsection{Optics of a monolayer.}
 Consider the graphene monolayer at $z=0$ with
 $\epsilon_0=\epsilon _g$
 deposited on the substrate ($z>0$) with the dielectric constant
$\epsilon_0=\epsilon _s$ (for graphene suspended in the vacuum,
$\epsilon_s=1$). The ac field is given by the sum of incident and
reflected waves in the vacuum, $z<0$, and by the transmitted wave
in the substrate. In the geometry considered, the current in
graphene monolayer can be written in the form
\begin{equation} j_x=\sigma(\omega)
\delta(z)E_x\, .\label{car} \end{equation}

Making use of the Fourier transformations with respect to the $x$
coordinate, ${\bf E}\propto e^{i k_xx}$,  we rewrite the Maxwell
equations (\ref{maxwell}) as follows
\begin{eqnarray}\label{mxeqn1}
\begin{array}{cc}
\displaystyle
ik_x\frac{dE_z}{dz}-\frac{d^2E_x}{dz^2}-\epsilon_0\frac{\omega^2}{c^2}E_x
=\frac{4\pi i\omega}{c^2}j_x \, ,&\\
\displaystyle
ik_x\frac{dE_x}{dz}+(k_x^2-\epsilon_0\frac{\omega^2}{c^2})E_z=0 \,
.&
\end{array}
\end{eqnarray}

The boundary conditions for these equations  at  $z=0$
   are  the continuity of the field component $E_x$ and the
jump of the electric-induction $z$ component   at the sides of the
monolayer:
 \begin{eqnarray}\label{jump}
\epsilon_sE_z|_{z=+0}- E_z|_{z=-0}=4\pi
\int_{-0}^{+0}\rho(\omega,k_x,z)dz\, .
\end{eqnarray}
The carrier density is connected to the current in Eq. (\ref{car})
according to the continuity equation
$$\rho(\omega,k_x,z)=j_x(\omega,k_x,z)k_x/\omega.$$
Substituting $E_z$ from the second Eq. $(\ref{mxeqn1})$ into
(\ref{jump}), we find the second boundary condition
\begin{equation}\frac{\epsilon_s}{k_s^2}\frac{dE_x}{dz}|_{z=+0}-
\frac{1}{(k_z^{i})^2}\frac{dE_x}{dz}|_{z=-0}=\frac{4\pi
\sigma(\omega)}{i\omega}E_x|_{z=0}\, , \label{bc}\end{equation}
 where
$$k_s=\sqrt{\epsilon_s(\omega/c)^2-k_x^2},\quad
k_z^i=\sqrt{(\omega/c)^2-k_x^2}\, .$$

Using the boundary conditions, we find the reflection (r) and
transmission (t) amplitudes
 \begin{equation}
 r =  \frac{1-C}{1+C}\,,\quad t=\frac{2}{1+C} ,
\label{tr}\end{equation} where
$C=k_z^i[(4\pi\sigma(\omega)/\omega)+(\epsilon_s/k_s)]\, .$ The
very simple results can be written for   graphene suspended in the
vacuum, when $\varepsilon_s=1$ and $k_s=k_z^i$. Then,  the
coefficient $C$ is close to unity,
\begin{equation}
C=1+4\pi\sigma(\omega)\cos\theta/c\, , \label{tras}\end{equation}
where $\theta$ is the incidence angle. Therefore, Eqs. (\ref{tr})
yield for the reflected and transmitted amplitudes
\begin{equation}
 r = -2\pi\sigma(\omega)\cos\theta/c \,,\quad t=1-2\pi\sigma(\omega)\cos\theta/c,
\label{trlim}\end{equation}
\begin{figure}[]
\noindent\centering{
\includegraphics[width=90mm]{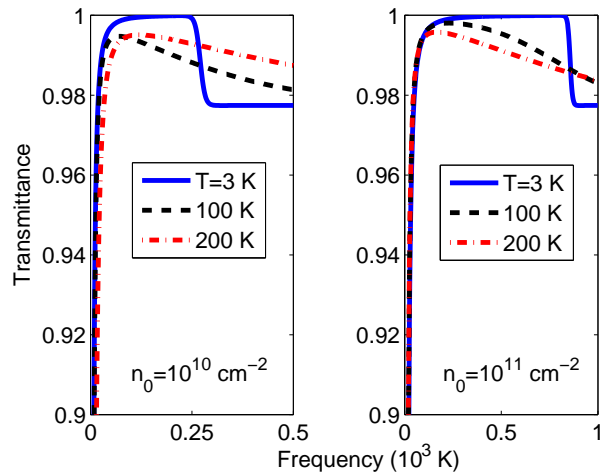}
} \caption{(color online). Transmittance spectrum of  graphene
with carrier densities $n_0=10^{10}$ cm$^{-2}$ (left) and
$n_0=10^{11}$ cm$^{-2}$ (right) versus the frequency  at
temperatures noted at the curves; normal incidence. For the
carrier density $n_0=10^{11}$ cm$^{-2}$, the chemical potential
equals to 428, 389, and 294 K at 3, 100, and 200 K,
correspondingly.} \label{figfr}
\end{figure}


The transmission coefficient $|t|^2$ calculated with the help of
Eqs. (\ref{sigm}),(\ref{equat6}),  and (\ref{tr}) for  normal
incidence is shown in Figs. \ref{figfr}
as a function of frequency, temperature, and carrier density. One
can see that the deviation of the graphene transmittance from
unity is proportional to the dimensionless parameter
$e^2max(T,\mu)/\hbar c\, \omega$, i.e.,  it is relatively large at
the very low frequencies $\omega\approx e^2max(T,\mu)/\hbar c$,
when the intraband conductivity plays the leading role.  At higher
frequencies, the interband transitions play the leading role.
Then,  the transmittance does not depend on frequency, being
controlled by the fine structure constant $e^2/\hbar c$. For
instance, at normal incidence and $T\ll\mu<\omega/2$, the
transmission coefficient is given by
\begin{equation}
  |t|^2=1-\frac{4\pi}{c} Re\,\sigma(\omega)=1-\pi\frac{e^2}{\hbar
  c}\,,
  \label{transmis}\end{equation} where the logarithmic terms is omitted since
  it contains the  fine structure
constant squared.  As was found recently \cite{NBG}, the linear
(in the fine structure constant)  effect can be observed in
graphene at the visible frequencies.

Reflectance increases with the temperature, because the carrier
density increases. As the chemical potential decreases with the
temperature,  the temperature dependence of reflectance is not
monotonic as seen clearly in Fig. \ref{figfr}.

\subsection{ Spectroscopy of graphene multilayers.}
Let the multilayers cross the $z$ axis at points $z_n=nd$, where
$d$ is the distance between the layers (see Fig. \ref{mu}b). Such
a system can be considered as a model of graphite since the
distance $d=3.35 \AA$ in graphite is larger than the interatomic
distance in the layer. So, we describe the carrier interaction in
the presence of  ac electric field with the help of
self-consistent Maxwell's equations (\ref{maxwell}). For the $x$
component of the field $E_x$, they give
\begin{equation}\label{max1}
 \left(\frac{d^2}{dz^2}+k_s^2+2k_s {\cal
D}\sum_{n}\delta(z-nd)\right) E_x=0\, ,
\end{equation}
where $ {\cal D}=2i\pi\sigma(\omega)k_s/\epsilon_g\omega\, . $

For the infinite number of layers in the stack, the solutions of
Eq. (\ref{max1}) represent two Bloch states
\begin{eqnarray}\nonumber
e_{1,2}(z)=e^{\pm ik_znd}\{\sin{k_s(z-nd)}- e^{\mp ik_zd}\times\\
\sin{k_s[z-(n+1)d]}\},\quad nd<z<(n+1)d \label{tsol}\end{eqnarray}
with  the quasi-momentum $k_z$ determined from
 the dispersion equation
\begin{equation}\label{dr}
\cos{k_zd}=\cos{k_s d}-{\cal D}\sin{k_s d}\, .
\end{equation}
The dispersion equation describes the electric field excitations
in the system. The quasi-momentum $k_z$ can be restricted to the
Brillouin half-zone $0<k_z<\pi/d$, if the parameter ${\cal D}$ is
real. In the general case, while taking
 the interband absorbtion into account,
 we fix the choice of the eigen-functions in
Eq.~(\ref{tsol}) by the condition ${\rm Im}\, {k_z}>0$ so that the
solution $e_{1}$ decreases in the positive direction $z$.

\begin{figure}[]
\noindent\centering{
\includegraphics[width=90mm]{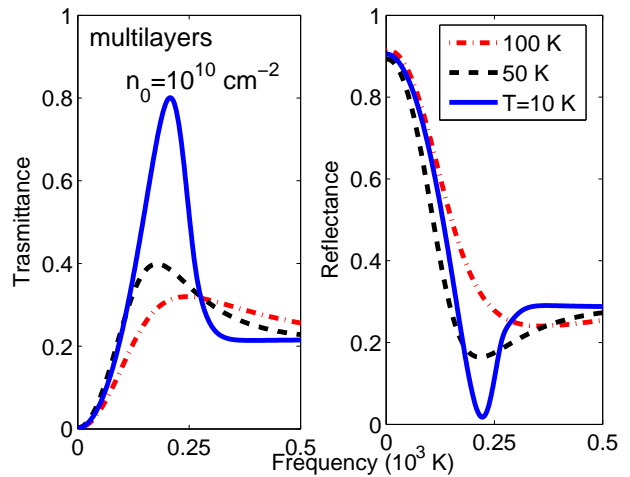}
} \caption{(color online). Transmittance and reflectance at the
normal incidence for a plate with the graphene multilayers
 and the carrier density $n_0=10^{10}$ cm$^{-2}$ in a layer; the
distance  between layers $d=3.35 \AA$, thickness of the plate is
$l=100 d$, temperatures are noted at curves.} \label{ref3-2d}
\end{figure}

In the long-wave limit, $k_z, k_s\ll 1/d$, the dispersion equation
(\ref{dr}) can be simplified. In this case, the dielectric
permittivity can be introduced not only in the normal $z$
 direction,
$\epsilon_{zz}=\epsilon_g\simeq 2.5$, but   in the tangential
direction as well, namely,
\begin{equation}
\epsilon_{xx}=\epsilon_g+4\pi i\sigma(\omega)/\omega d\,,
\label{exx}\end{equation} where $\sigma(\omega)$ is the dynamic
conductivity, Eq. (\ref{sigma}), of the one layer.
 Then the dispersion equation
(\ref{dr}) takes the form
\begin{equation}\label{dumac}
k_x^2\epsilon_{xx}+k_z^2\epsilon_{zz}=
(\omega/c)^2\epsilon_{xx}\epsilon_{zz}\,.
\end{equation}
We see from this equation, that the weakly damped solutions (for
normal incidence, $k_x=0$) exist, if the real part of the
dielectric function $\epsilon_{xx}$ is positive and larger than
the imaginary part.  According Eqs. (\ref{si}) and (\ref{ibd}),
this condition is fulfilled just below the threshold of interband
transitions. Due to these weakly damped waves, the maximum in
transmittance and the corresponding minimum in reflectance of the
multilayer system appear at  temperatures below   the threshold as
shown in Fig. \ref{ref3-2d}.


Calculations done for a thin plate with  the stack of   graphene
multilayers give the following result for the transmission
amplitude
\begin{equation}
t=\frac{4k_zk_z^i}{(k_z+k_z^i)^2f-(k_z-k_z^i)^2f^{-1}}\,,
\label{trpl}\end{equation} where $f=\exp(-ik_zl),$
$k_z^i=(\omega/c)\cos{\theta}$, $k_z$ is  determined by the
  dispersion equation (\ref{dumac})
 at  fixed values of $\omega$ and $k_x$.

At higher frequencies,  $\omega>2\mu=270$ K (in Fig.
\ref{ref3-2d}), after the downfall, the reflection coefficient
corresponds with the interband absorption presented by the
$\theta-$singularity in the real part of conductivity.  In
contrast to the monolayer, the effect of carriers in the
multilayers is controlled by the large dimensionless factor
$(e^2/\hbar \epsilon_g\omega d)$ [see Eq.( \ref{exx})] of the
function varying rapidly at $\omega\approx 2\mu$. It means that
the interband absorption can be observed at relatively large
frequencies  and low temperatures. Notice that observations of the
absorbtion threshold provide a direct method of carrier density
 characterization  in graphene.

\section{Conclusion}
In conclusion, we have developed the detailed microscopic theory
of   the graphene monolayer and multilayers spectroscopy. We have
shown that the  reflectance from the monolayer is determined for
infra-red region by the intraband Drude-Boltzmann conductivity and
for  higher frequencies by the interband absorption. We have
argued that at low temperatures and high carrier densities,
 the reflectance from multilayers  has the sharp
downfall with the subsequent plateau. These features are caused by
excitations of propagating waves and  the direct interband
electron transitions.

{\bf Acknowledgments}\\[0.1cm]
 This work is supported by the Russian Foundation for Basic
Research (grant No. 07-02-00571).


\begin{thebibliography}{99}
\bibitem{FV} L.A. Falkovsky and A.A. Varlamov, cond-mat/0606800,
Eur. Phys. J. B \textbf{56}, 281 (2007).

\bibitem{GSC} V.P. Gusynin, S.G. Sharapov, and J.P.Carbotte,
 Phys. Rev. \textbf{B 75}, 165407 (2007);
 cond-mat/0607727, Phys. Rev. Lett. \textbf{96}, 256802 (2006).

\bibitem{Cs} J. Cserti, Phys. Rev. \textbf{B 75}, 033405 (2007).


\bibitem{FP} L.A. Falkovsky and S.S. Pershoguba,
Phys. Rev.  {\bf B 76}, 153410 (2007).

\bibitem{ARF} D.S.L. Abergel, A. Russsell, and V. I. Fal'ko,
cond-mat/0705.0091, unpublished.

\bibitem{NBG} R.R. Nair, P. Blake, A.N. Grigorenko, K.S. Novoselov,
T.J. Blooth, T. Stauber, N.M.R. Peres, and A.K. Geim,
cond-mat/0803.3718, unpublished.







\end{thebibliography}
\end{document}